\documentclass[12pt,preprint]{aastex}

\slugcomment{Accepted for publication in ApJ}

\shorttitle{The orbit of the brown dwarf binary GJ\,569B}
\shortauthors{Lane et al$.$}

\begin{document}

\title{ The Orbit of the Brown Dwarf Binary Gl\,569B }

\author{B.\,F. Lane, M.\,R. Zapatero Osorio, M.\,C. Britton}
\affil{Department of Geological \& Planetary Sciences, MS 150-21, 
       California Institute of Technology, Pasadena CA 91125, U.S.A.}
\email{ben,mosorio,mbritton@gps.caltech.edu}

\author{E.\,L. Mart\'\i n}
\affil{Institute for Astronomy, University of Hawaii at Manoa, 2680 
       Woodlawn Drive, Honolulu HI 96822, U.S.A.}
\email{ege@IfA.Hawaii.Edu}

\and

\author{S.\,R. Kulkarni\altaffilmark{1}}
\affil{Department of Geological \& Planetary Sciences, MS 150-21, 
       California Institute of Technology, Pasadena CA 91125, U.S.A.}
\email{srk@astro.caltech.edu}

\altaffiltext{1}{Department of Astronomy, California Institute of 
                 Technology, Pasadena CA 91125, U.S.A. }

\begin{abstract}
  We present photometric, astrometric and spectroscopic observations
  of the nearby (9.8\,pc) low-mass binary Gl\,569Bab (in turn being a
  companion to the early-M star Gl\,569A), made with the Keck adaptive
  optics facility. Having observed Gl\,569Bab since August 1999, we
  are able to see orbital motion and to determine the orbital
  parameters of the pair. We find the orbital period to be $892 \pm
  25$ days, the semi-major axis to be $0.90 \pm 0.02$ AU, the
  eccentricity to be $0.32 \pm 0.02$ and the inclination of the system
  to be $34 \pm 3$ degrees (1-$\sigma$).  The total mass is found to
  be $0.123_{-0.022}^{+0.027} M_{\odot}$ (3-$\sigma$). In addition, we
  have obtained low resolution ($R$\,=\,1500--1700) near-infrared
  spectra of each of the components in the $J$- and $K$-bands. We
  determine the spectral types of the objects to be M8.5V (Gl\,569Ba)
  and M9V (Gl\,569Bb) with an uncertainty of half a subclass. We also
  present new $J$- and $K$-band photometry which allows us to
  accurately place the objects in the HR diagram. Most likely the
  binary system is comprised of two brown dwarfs with a mass ratio of
  0.89 and with an age of approximately 300\,Myr.

\end{abstract}


\clearpage
\section{Introduction}                  

Brown dwarfs (BDs), despite sometimes being labeled ``failed stars'',
are very interesting objects. According to our current understanding,
BDs may represent the extreme low-mass end of star formation in which
the mass is too small to sustain thermonuclear fusion. This
stellar-substellar transition (also defined as the minimum mass at
which the internal energy provided by nuclear burning quickly balances
the gravitational contraction energy) is expected to occur around
0.075\,$M_\odot$ (Baraffe et al$.$ 1998)\nocite{baraffe98} for objects
with solar metallicities. This value is slightly lower in recent
models, which predict a transition mass of 0.072\,$M_\odot$ (Chabrier
\& Baraffe 2000\nocite{chabrier00}).  Objects with masses below this
limit never reach the stable hydrogen burning main sequence, but
instead cool down as they age (Burrows et al$.$
1997\nocite{burrows97}; Chabrier et al$.$ 2000\nocite{chabrier00a}) so
that their surface temperatures and luminosities strongly depend on
age as well as mass. As these cooling curves rely on poorly tested
theoretical models, it is highly desirable to calibrate them with
direct measurements.

Obtaining dynamical masses for very low-mass (VLM,
$M\,\leq\,0.2\,M_{\odot}$) stars and BDs from binaries is a
challenging prospect, as so far there are no known eclipsing binary
VLM stars or BDs. However, there are a number of known wide,
non-eclipsing VLM and BD binaries (Kirkpatrick et al$.$
2001\nocite{kirk01}; Gizis, Kirkpatrick \& Wilson
2001\nocite{gizis01}; Leinert et al$.$ 2001\nocite{leinert01}; see
also Reid et al$.$ 2001a\nocite{reid01a} and references therein) that
have been observed with a range of instruments and techniques,
including ground-based infrared and optical imaging, speckle
interferometry, adaptive optics (AO), and the Hubble Space Telescope.
These binary systems promise to yield highly accurate dynamical
masses, although they tend to have long periods and hence will require
patience.

In the absence of direct mass measurements for VLM/BD objects, one has
to rely on indirect methods that may constrain mass ranges but that do
not provide high-precision mass values. One very useful such technique
is the lithium test (Magazz\`u, Mart\'\i n \& Rebolo
1993)\nocite{magazzu93}.  Lithium is an element that is easily
destroyed under the conditions prevalent in stellar interiors at
temperatures slightly below those required for hydrogen burning.
Objects more massive than $\sim0.060\,M_\odot$ have their primordial
lithium abundances depleted as long as they are fully convective. The
lithium test has been used to confirm BD candidates in the Pleiades
(e.g., Basri, Marcy \& Graham 1996)\nocite{basri96} and in the field
(e.g., Mart\'\i n, Basri \& Zapatero Osorio 1999\nocite{martin99};
Kirkpatrick et al$.$ 1999\nocite{kirk99}; Tinney, Delfosse \&
Forveille 1997\nocite{tinney97}).

Gl\,569A is a nearby ($d$\,=\,9.8\,pc), chromospherically active
late-type (M2.5V) star.  Forrest, Skrutskie \& Shure
(1988)\nocite{forrest88} first reported a possible BD companion with a
separation of 5\arcsec~from the primary. Based on a low resolution
spectrum Henry \& Kirkpatrick (1990)\nocite{henry90} classified this
object as an M8.5 dwarf with a mass of $0.09 \pm 0.02 M_\odot$.  More
recently, Mart\'\i n et al$.$ (2000a)\nocite{martin00a} resolved the
companion into two separate objects (Gl\,569Ba and Bb) using AO
observations with the Keck\,II telescope.  They estimated the orbital
period of the Ba-Bb binary to be around 3 years, and the total mass of
the binary pair to be in the range $0.09-0.15\,M_\odot$. Herein we
present the results of extensive follow-up observations of this
interesting pair, including improved photometry as well as near-IR
spectroscopy and astrometry of the resolved binary components. We use
the photometry to accurately place the objects in the
Hertzsprung-Russell (HR) diagram, the spectroscopy to derive spectral
types, and the astrometry to derive the total mass of the system from
its orbital parameters.

\section{NIR photometry and astrometry}
\subsection{The composite pair: Gl\,569Bab}

Broad-band near-infrared photometry of the composite pair Gl\,569Bab
is available in the literature (Forrest et al$.$
1988\nocite{forrest88}; Becklin \& Zuckerman 1988\nocite{becklin88}).
However, the measurement uncertainties claimed by the authors are too
large for an accurate placement of these objects in the HR diagram or
for direct comparison with theoretical evolutionary models. With the
objective of improving the photometric data, we have collected $J$ and
$K_{short}$ direct images of the system Gl\,569A and Gl\,569Bab with
the near-infrared camera (Hg\,Cd\,Te detector, 256\,$\times$\,256
elements) mounted at the Cassegrain focus of the 1.5-m Carlos
S\'anchez Telescope (CST, Teide Observatory) on 2001 February 8. The
observations were performed through the ``narrow-optics'' of the
instrument, which provides a pixel projection of 0.4\arcsec~onto the
sky.  The atmospheric seeing conditions during the night of the
observations were fairly stable around 1\arcsec, which allowed us to
easily separate the M2.5-type star from the pair Gl\,569Bab. This
latter object was not resolved into its two components. The total
integration times were 5\,s and 40\,s in $J$ and $K_{short}$ filters,
respectively. A five-position dither pattern was used to obtain the
images; each image consisted of 4 ($J$) or 8 ($K$) co-added exposures
of 0.25\,s ($J$) and 0.5\,s ($K$) respectively. The dither pattern was
repeated twice for the K-band observations.
 
Dithered images were combined in order to obtain the sky background,
which was later substracted from each single frame. Gl\,569Bab is
clearly detected in individual images, and we have obtained aperture
photometry on each of them using PHOT in IRAF.\footnote{IRAF is
distributed by National Optical Astronomy Observatories, which is
operated by the Association of Universities for Research in Astronomy,
Inc., under contract with the National Science Foundation.}
Instrumental magnitudes were placed on the UKIRT photometric system
using observations of the standard star HD\,136754 (Casali \& Hawarden
1992\nocite{casali92}), which was imaged with the same instrumental
configuration just before and immediately after our target. Both the
science target and the standard star were observed at similar air
masses. The photometric error of the calibration was $\pm$0.03\,mag in
both filters. $K_{short}$ displays a different bandpass compared to
$K_{UKIRT}$; the transformation between these two filters is not well
defined yet, albeit for objects as red as Gl\,569Bab it has been
estimated at $K_{short}$\,--\,$K_{UKIRT}$\,=\,0.035 (Hodgkin et al$.$
1999\nocite{hodgkin99}, and references therein). We have applied this
correction to our photometry as well as the relations given in
Leggett, Allard \& Hauschildt (1998)\nocite{leggett98} to convert
UKIRT data into the CIT photometric system. The final average
magnitudes derived for Gl\,569Bab are given in Table~\ref{phot}, where
the photometric errors listed correspond to typical 1\,$\sigma$
uncertainties of single measurements.

The astrometry of Gl\,569Bab relative to the bright primary Gl\,569A
as measured on the CST data (MJD\,=\,51948.202) is the following:
angular separation of 4.\arcsec890\,$\pm$\,0.\arcsec040, and position
angle of 30$^{\circ}$\,$\pm$\,3$^{\circ}$. We note that these values
differ from those published in Forrest et al$.$ (1988)\nocite{forrest88}
by more than 2\,$\sigma$, providing evidence of the orbital motion of
Gl\,569Bab around the M2.5-type star in the time interval of roughly
15\,years.

\subsection{Adaptive Optics Imaging of Gl\,569Ba-Bb pair}

\begin{figure}[h]
\plotone{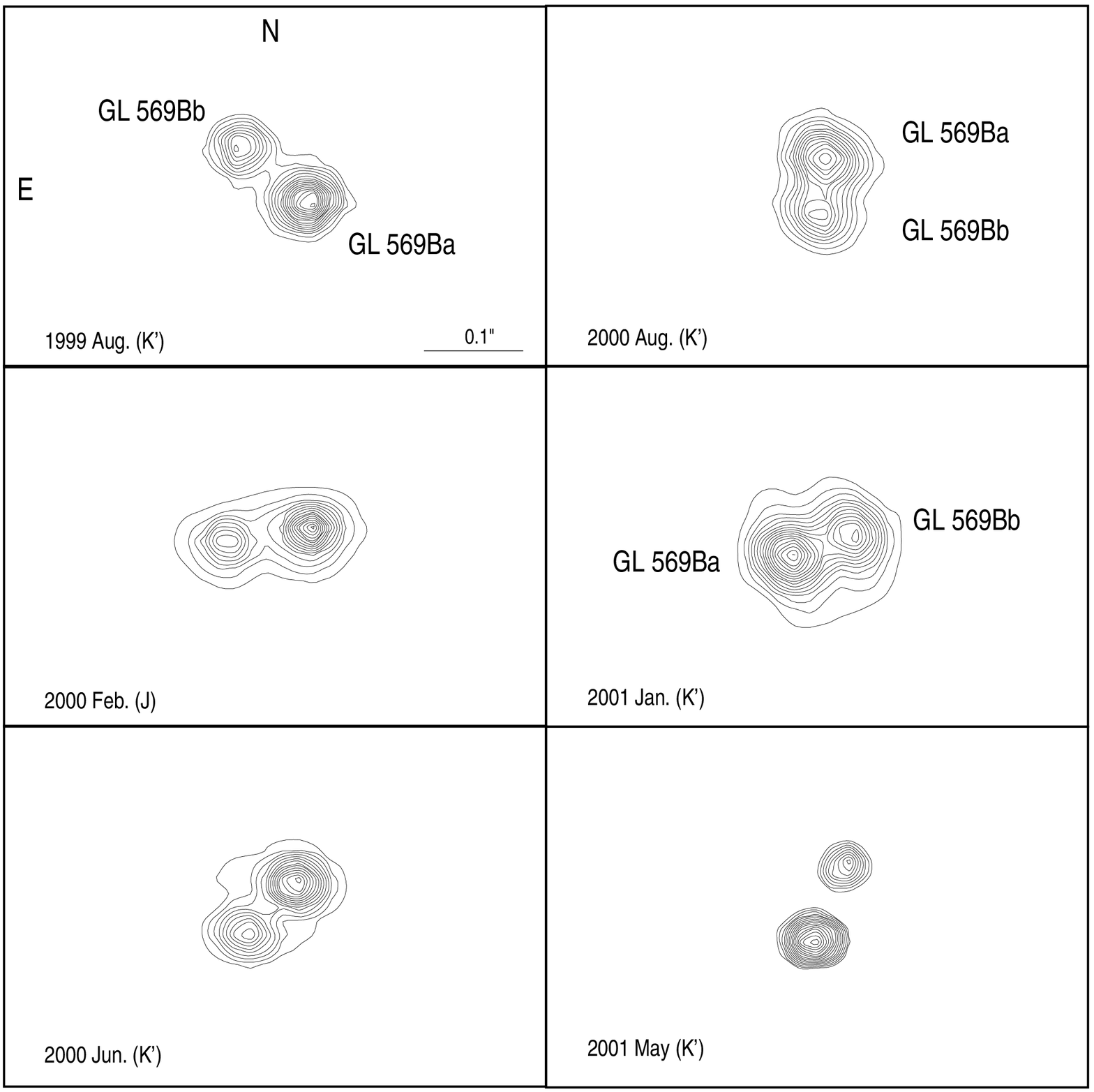}
\caption[]{\label{astro} Contour images of the Gl\,569Ba-Bb pair (near-IR filter is given in brackets) showing the orbital motion. These data have been obtained with the Adaptive Optics facility of the Keck\,II telescope and with the KCAM (first 2 epochs) and SCAM/NIRSPEC (last 5 epochs) instruments. Data of 1999 August, when the binary was resolved for the first time, were presented in Mart\'\i n et al$.$ (2000a)\nocite{martin00a}.}
\end{figure}

Gl\,569Bab was observed on 7 occasions between 1999 August and 2001
May with the Keck II AO system (Wizinowich et al$.$
1988)\nocite{wizin98}. The first 3 observations made use of the KCAM
camera with a NICMOS-3 infrared array. The later observations made use
of the slit-viewing camera (SCAM) associated with the NIRSPEC
instrument (McLean et al$.$ 1998\nocite{mclean98}).  SCAM uses a
PICNIC Hg\,Cd\,Te array. For both cameras the pixel scale was
0.0175\arcsec~and the field of view was
$4.48\arcsec\,\times\,4.48$\arcsec, except for the 2001 May
observation when the SCAM pixel scale was changed to 0.0168\arcsec.
Exposures were generally obtained in the $K'$-band, except for 2000
Febuary when the observations were taken in the $J$-band. Exact
exposure times varied, but typically consisted of 30 coadded 2-second
exposures. We used a 1\%~transmission neutral density filter in the
beam to prevent saturating the bright primary star. Flat-field
correction was performed using twilight exposures, while sky
subtraction made use of images of an adjacent field observed
immediately after Gl\,569B.  Corrected seeing varied between
0.05\arcsec~and 0.08\arcsec. For three of the observations the primary
star (Gl\,569A) was also in the field of view, providing an in-field
astrometric reference.  For the other observations the primary was
either not observed or was saturated. No photometric standard stars
were observed in any of the epochs, so we cannot provide absolute
photometric calibrations for the AO observations.  Figure~\ref{astro}
shows the resulting images of the pair Gl\,569Bab at six different
epochs. This figure clearly demonstrates that this system is resolved
into a binary and that orbital motion is evident.

We used the DAOPHOT package (in IRAF) for data reduction and analysis.
The point spread function of the objects in each frame were fitted
with an elliptical Gaussian function, providing relative astrometry
(Table \ref{ast}) and photometry.  The relative photometry of
Gl\,569Ba and Gl\,569Bb was derived by computing the ratio of the
amplitudes of the best fitting Gaussians (Table \ref{phot}). Gl\,569Bb
is fainter by 0.51\,$\pm$\,0.02\,mag and 0.41\,$\pm$\,0.03\,mag in the
$J$- and $K$-bands, respectively. This makes this object redder in
($J-K$) by 0.10\,$\pm$\,0.04\,mag. The error bars take into account
the dispersion observed from image to image, and from one observing
run to another. We do not find a significant relative photometric
variability in any member of the pair within 3\,$\sigma$ the
uncertainties. With the relative brightness of the two components and
the combined flux known, it is possible to derive the individual
absolute magnitudes of Gl\,569Ba and Gl\,569Bb. We list in
Table~\ref{phot} the resulting decomposition for the $J$- and
$K$-bands.  The corresponding error bars incorporate the photometric
uncertainties of the combined system Gl\,569Bab and the uncertainties
of the relative photometry. We are confident that the latter is
determined with a higher accuracy.

\section{Low resolution NIR spectra}
We have obtained low resolution spectra of Gl\,569Ba and Gl\,569Bb in
the $J$- (1.158--1.368\,$\mu$m) and $K$-bands (1.992--2.420\,$\mu$m)
using the cross-dispersion spectrograph NIRSPEC and the AO facility at
the Keck\,II telescope. The data were collected on 2000 June 20. The
raw seeing and transparency conditions were very good during the
observations, and the AO correction applied to the primary star
Gl\,569A provided well resolved images of the binary system Gl\,569Bab
(AO corrected seeing of 0.05\arcsec). NIRSPEC in its spectroscopic
mode is equipped with an Aladdin InSb 1024\,$\times$\,1024 detector
with a pixel projecting 0.0185\arcsec~onto the sky. For the present
study, we selected the low resolution spectroscopic mode which
provides nominal dispersions of 2.8\,\AA/pix and 4.2\,\AA/pix in the
$J$- and $K$-bands, respectively. The 3\,pixel-wide slit was aligned
with the two components of the binary (PA\,$\sim$\,139\,deg) so that
both targets were observed simultaneously.

Total exposure times were 240\,sec and 400\,sec for the $J$-band and
$K$-band spectra, respectively. The observing strategy employed was as
follows: 6 ($J$) and 10 ($K$) individual integrations of 20\,sec ($J$)
and 10\,sec ($K$) each at two different positions along the entrance
slit separated by about 1.8\arcsec. This procedure was repeated twice
in the $K$-band. In order to remove telluric absorptions due to the
Earth's atmosphere, the near-infrared featureless A0V-type star
HR\,5567 was observed very close in time and in air mass (within 0.05
air masses). Calibration images (argon arc lamp emission spectra and
white-light spectra) were systematically taken after observing each
source.

Raw data were reduced following conventional techniques in the
near-infrared. Nodded images were subtracted to remove the sky
background and dark current. The spectra of the sources and of the
calibration lamps were then extracted using subroutines of the
TWODSPEC package available in IRAF. The extraction apertures of
Gl\,569Ba and Gl\,569Bb were selected so that cross-contamination was
less than 10\%. The extracted spectra of the sources were divided by
their corresponding normalized extracted flat-fields, and calibrated
in wavelength. The 1\,$\sigma$ dispersion of the fourth-order
polynomial fit was 0.4\,\AA~and 1.0\,\AA~in the $J$ and $K$ spectra,
respectively. The hydrogen P$\beta$ absorption line at 1.2818\,$\mu$m
and the B$\gamma$ absorption line at 2.1655\,$\mu$m in the spectra of
HR\,5567 were interpolated before they were used for division into the
corresponding science spectra. We are confident that the science
spectra have good cancellation of atmospheric features. To complete
the data reduction, we multiplied the spectra of our targets by the
black body spectrum for the temperature of 9480\,K, which corresponds
to the A0V class (Allen 2000\nocite{allen00}).

\subsection{Spectral types, atomic and molecular features}
\begin{figure}[h]
\epsscale{0.5}
\plotone{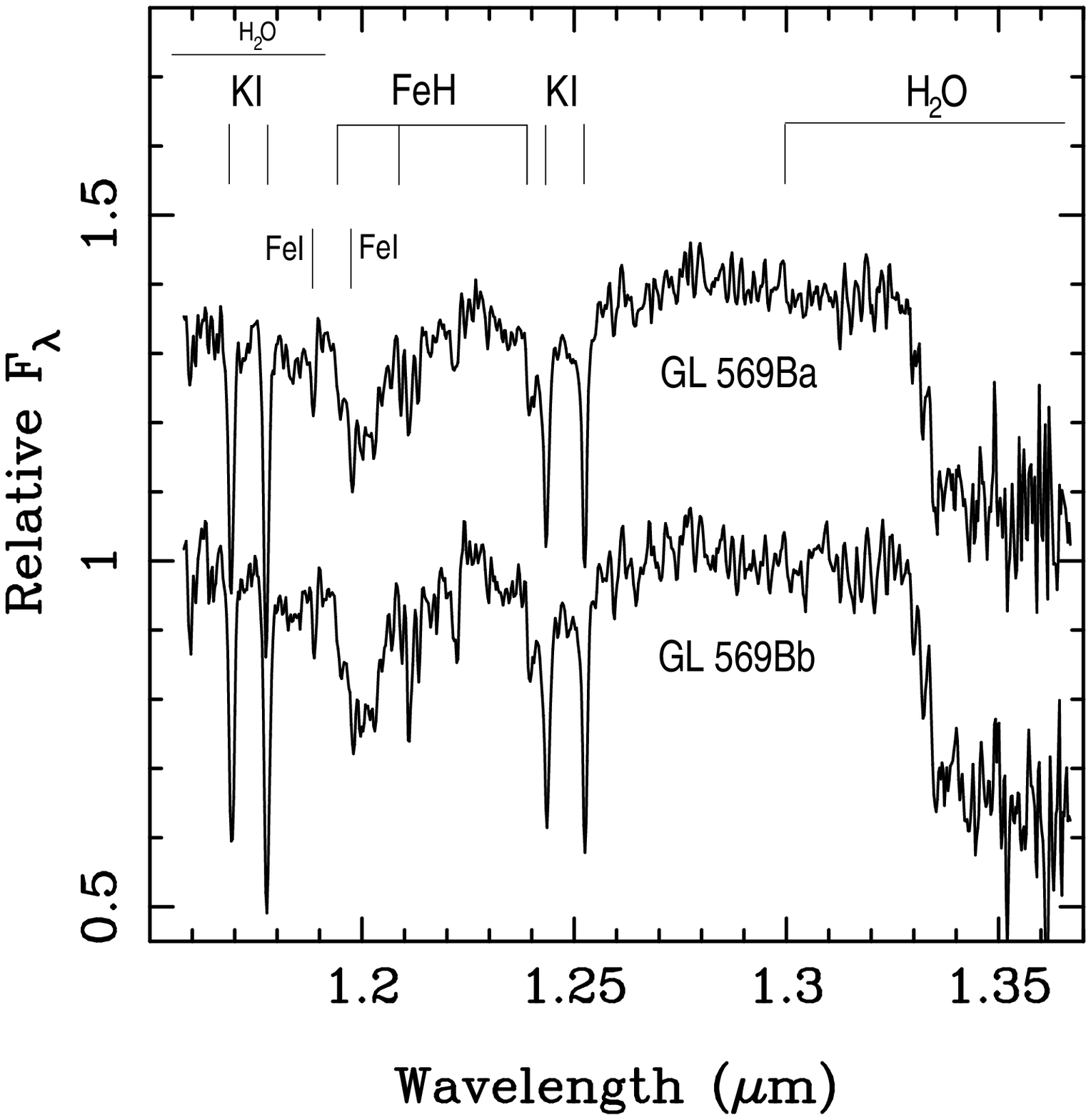}
\plotone{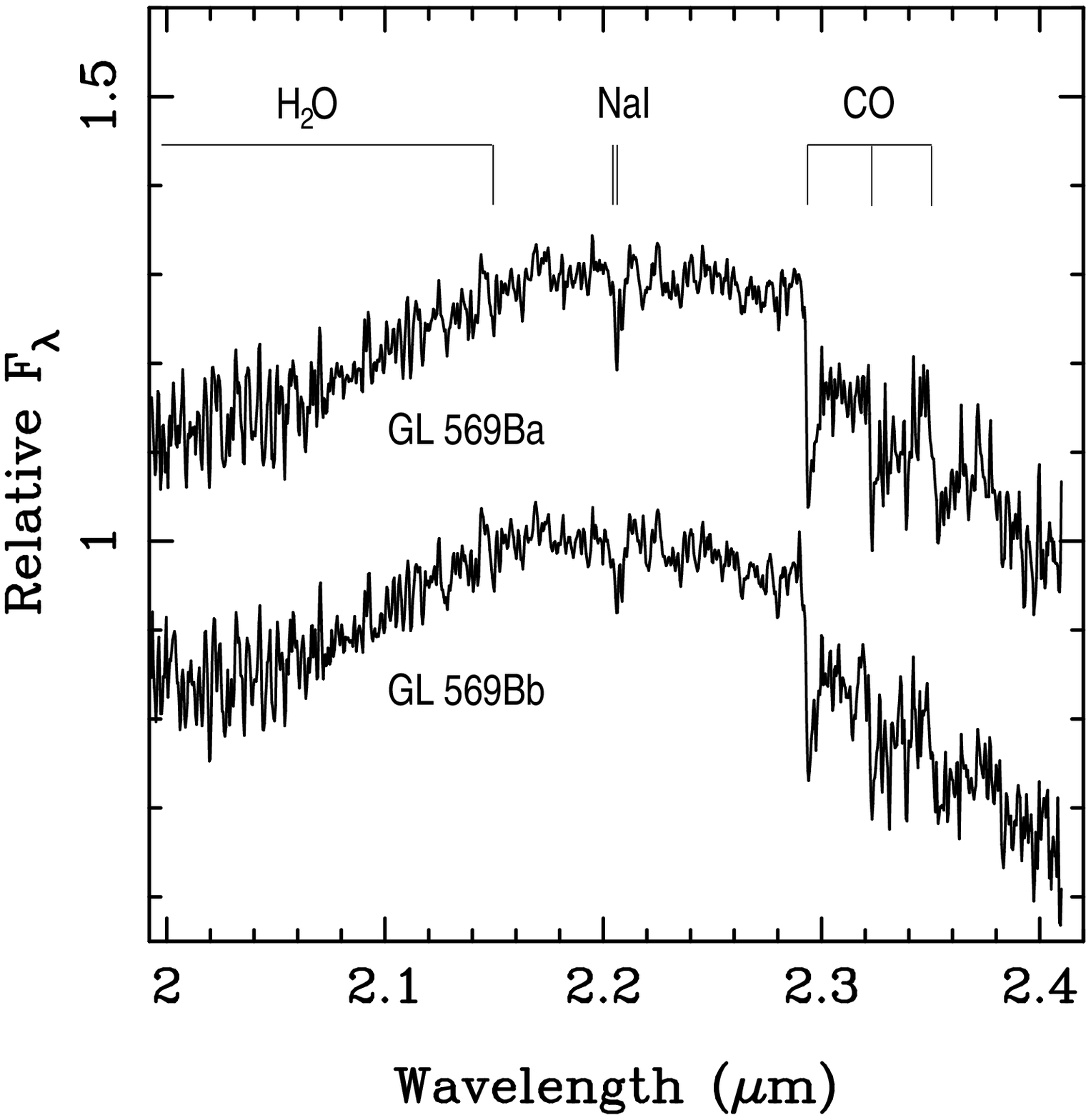}
\caption[]{\label{spec} $J$-band (top panel) and $K$-band (lower
  panel) NIRSPEC spectra of Gl\,569Ba and Gl\,569Bb obtained using the
  AO system of the Keck\,II telescope. Some features have been
  identified after Jones et al$.$ (1994) and McLean et al$.$ (2000).
  The spectra have been normalized to unity at 1.29\,$\mu$m and at
  2.19\,$\mu$m. An offset has been added to Gl\,569Ba's data for
  clarity. }
\end{figure}

The resultant average spectra with a resolution of $R$\,=\,1500 in $J$
and $R$\,=\,1700 in $K$ are depicted in Fig.~\ref{spec}. The strongest
molecular and atomic features are indicated following the
identifications provided by Jones et al$.$ (1996)\nocite{jones96} and
McLean et al$.$ (2000)\nocite{mclean00}. The spectra of both
components, Gl\,569Ba and Gl\,569Bb, are indeed very similar. The
composite spectrum of Gl\,569Bab in the optical has been previously
studied by Henry \& Kirkpatrick (1990)\nocite{henry90} and
Kirkpatrick, Henry \& McCarthy (1991)\nocite{kirk91}, who derived a
dwarf spectral type of M8.5. In addition, this object is listed in the
Table~1 of Kirkpatrick et al$.$ (1991)\nocite{kirk91} as a primary
dwarf spectral standard. Our data agree with this measurement for the
bright component Gl\,569Ba, and also provide evidence that Gl\,569Bb
is not significantly cooler. This is fully consistent with the
photometry presented above.

We have obtained equivalent widths of the strongest observed atomic
absorptions of K\,{\sc i} and Na\,{\sc i} in the spectra; the
measurements are given in Table~\ref{eqw}. Due to the low resolution
of our data, the majority of these lines are considerably blended with
other spectral features, e.g. the K\,{\sc i} line at 1.2435\,$\mu$m is
contaminated by a strong molecular band of FeH. The values in
Table~\ref{eqw} have been extracted adopting the base of each line as
the continuum. We find typical standard deviations in equivalent width
close to 10\%~over the reasonable range of possible continua. Although
this procedure does not give an absolute equivalent width, it is
commonly used by different authors, and allows us to compare our
values with those published in the literature. We have also measured
the strengths of the H$_2$O band at 1.330\,$\mu$m and the CO band at
2.294\,$\mu$m in a way similar to that described in McLean et al$.$
(2000)\nocite{mclean00}, Reid et al$.$ (2001b)\nocite{reid01b} and
Jones et al$.$ (1994)\nocite{jones94}. Our measurements are listed in
Table~\ref{eqw} with uncertainties of about 5\%. All these values are
comparable to those obtained from similar spectral type field stars,
which suggests that neither Gl\,569Ba nor Gl\,569Bb have very
discrepant metallicities or gravity.

We note that the equivalent widths of the K\,{\sc i} lines and the
H$_2$O and CO absorptions in Gl\,569Bb appear to be slightly larger
than in Gl\,569Ba, while the Na\,{\sc i} lines in the $K$-band
spectrum are smaller. This trend is observed for decreasing
temperatures (Jones et al$.$ 1994), and clearly indicates the cooler
nature of Gl\,569Bb. By fitting a polynomial spectral type-equivalent
width relation to the data available for spectral standard stars (see
Reid et al$.$ 2001b), we conclude that the differences between Ba \& Bb
in our measurements are consistent with Gl\,569Bb being half a subclass
cooler. This would make Gl\,569Bb an M9-dwarf ($\pm$0.5 subclasses).

\subsection{Radial velocities}
We used our low resolution near-IR spectra taken on 2000 June 20
(MJD\,=\,51715.365) to compute the relative radial velocity of
Gl\,569Bb and Gl\,569Ba via Fourier cross-correlation. Because the
spectroscopic data have been corrected for telluric lines, we do not
expect these lines to be a large source of uncertainty. Unfortunately,
no spectra were taken of the primary star Gl\,569A, so we cannot
determine the relative radial velocity of the pair with respect it.
The velocity dispersion of the data is rather poor
(1\,pixel\,$\sim$\,66\,km\,s$^{-1}$ in $J$ and
$\sim$\,57\,km\,s$^{-1}$ in $K$). Nevertheless, the cross-correlation
technique was able to achieve precisions of about 1/4\,pixel, so we
obtained a relative radial velocity accurate to about
15\,km\,s$^{-1}$. We verified this by cross-correlating individual
spectra of each component against itself. The relative velocity
(Gl\,569Bb cross-correlated with Gl\,569Ba) we measure is
25\,km\,s$^{-1}$ in $J$ and 6\,km\,s$^{-1}$ in $K$, with an average
value of 15.5\,km\,s$^{-1}$. The peak-to-peak radial velocity
variation of the system on the basis of the orbital solution presented
in next section is around 14\,km\,s$^{-1}$; our measurement is
consistent within the error bar with the expected value at the epoch
of the observations.  However, this error bar is rather large and
prevents us from making further analysis (like the presence of
invisible companions).  The maximum peak of the cross-correlation
function is in the range 0.93--0.97, which indicates the similarity
and the high signal-to-noise ratio of the spectra.

\section{Orbit determination and total mass of the pair}
\begin{figure}[h]
\epsscale{1.0}
\plotone{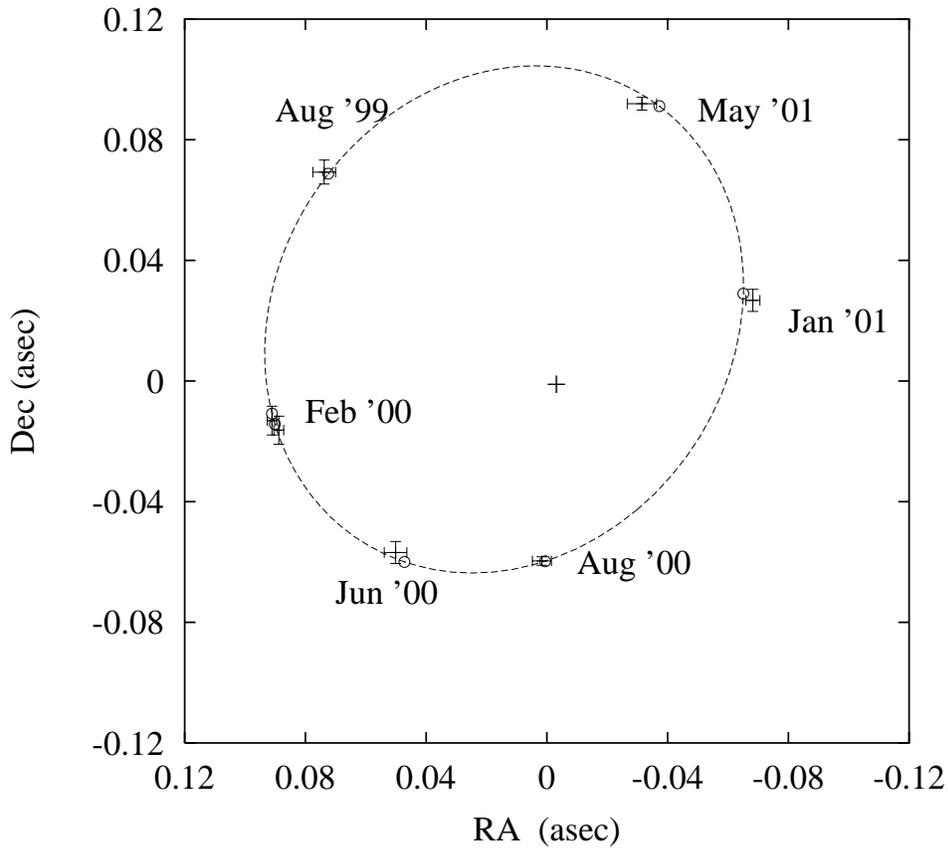}
\caption[]{\label{orbit} The relative astrometry of the Gl\,569Ba-Bb pair,
  together with the best-fit orbit (dotted ellipse). Error-bar crosses denote
  measurements and circles indicate the predicted location on the
  orbit at the time of the observations. North is up and East is to the left.}
\end{figure}

\begin{figure}[h]
\epsscale{1.0}
\plotone{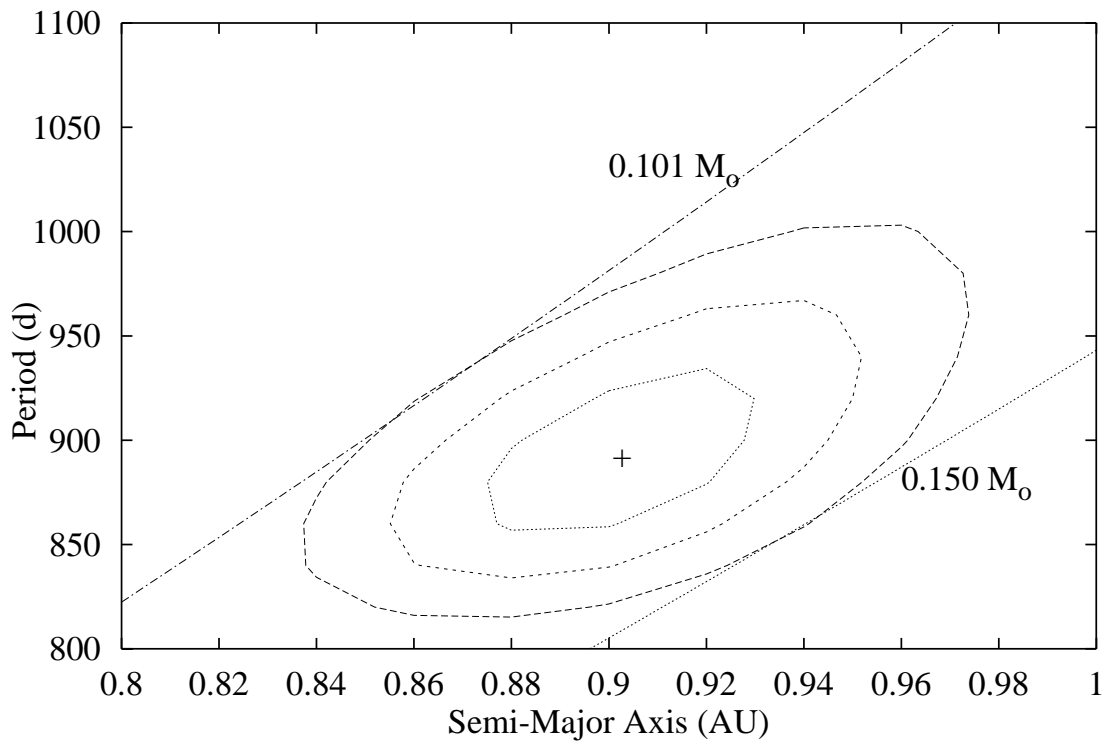}
\caption[]{\label{chi} The $\chi^2$ as a function of period and
semi-major axis. All other parameters are selected to provide the
lowest $\chi^2$.  The contours give the 1\,$\sigma$, 2\,$\sigma$ and
3\,$\sigma$ uncertainties in the two parameters. The two diagonal
lines correspond to combinations of period and semi-major axis giving
a total mass of 0.150\,$M_{\odot}$ and 0.101\,$M_{\odot}$,
respectively. These are the upper and lower limits that we have
adopted for the total mass of the binary Gl\,569Bab. The cross
indicates the preferred solution.}
\end{figure}

We determined the apparent orbit of the binary pair Gl\,569Bab by
fitting a Keplerian model to the relative astrometric data shown in
Table~\ref{ast}. As the fit is non-linear in the orbital elements, we
made use of a gradient-following fitting routine (Press
1992)\nocite{press92} to find the optimal (in a chi-squared, $\chi^2$,
sense) orbital parameter set. As there may be many local minima in the
$\chi^2$ manifold, and our gradient following routine moves strictly
downhill, we ensured that it found the global minimum by starting it
at a range of different locations in parameter space.  The best-fit
parameters are given in Table~\ref{params}, and the visual orbit is
shown together with the astrometry in Fig.~\ref{orbit}. The $rms$
of the residuals is 0.0024\arcsec, and the residuals do not show any
long-term drifts. Uncertainties (assuming normal errors) were
estimated from the covariance matrix of the fit, scaled by the reduced
chi-squared (0.96). However, as apparent in Fig.~\ref{chi} the
best-fit values for the semi-major axis and period are correlated,
leading to a lower uncertainty in the total mass. The mass uncertainty
was estimated by finding contours in parameter space where the
$\chi^2$ was increased by 2.3, 6.2 and 11.8 respectively,
corresponding to the 1\,$\sigma$, 2\,$\sigma$ and 3\,$\sigma$
contours. The 1\,$\sigma$, 2\,$\sigma$ and 3\,$\sigma$ mass ranges are
0.114--0.135, 0.107--0.142 and 0.101--0.150 $M_{\odot}$ respectively,
while the best-fit values for the period and semi-major axis
correspond to a total mass of 0.123 $M_{\odot}$.  From the relative
photometry we infer that the mass ratio of the binary is close to, but
not exactly, equal. Hence the 2\,$\sigma$ upper mass limit of the
secondary Gl\,569Bb is less than 0.071\,$M_{\odot}$, i.e$.$ below the
hydrogen burning mass threshold, making it a likely brown dwarf.

It is important to note that although we were able to obtain reliable
relative astrometry of the Ba-Bb pair (over a separation of $\sim
0.1$\arcsec), the uncertainties in plate scale and orientation and the
saturation of the bright star in some images were such that we were
unable to reliably measure the orbital motion of each component of the
Ba-Bb pair with respect to a separate reference, i.e$.$ Gl\,569A
(located $\sim5$\arcsec~away). Hence, while we can determine the
relative orbit of the Ba-Bb pair to a high degree of precision, we
cannot astrometrically determine the mass ratio of the two components.
Further studies are needed to confirm or discard the
substellar nature of the primary Gl\,569Ba. We will combine our
photometry (absolute and relative), the total mass of the system and
additional information available in the literature to compare the
binary with the most recent evolutionary models.

\section{Discussion}

\subsection{Color-magnitude diagram}
\begin{figure}
\epsscale{1.0}
\plotone{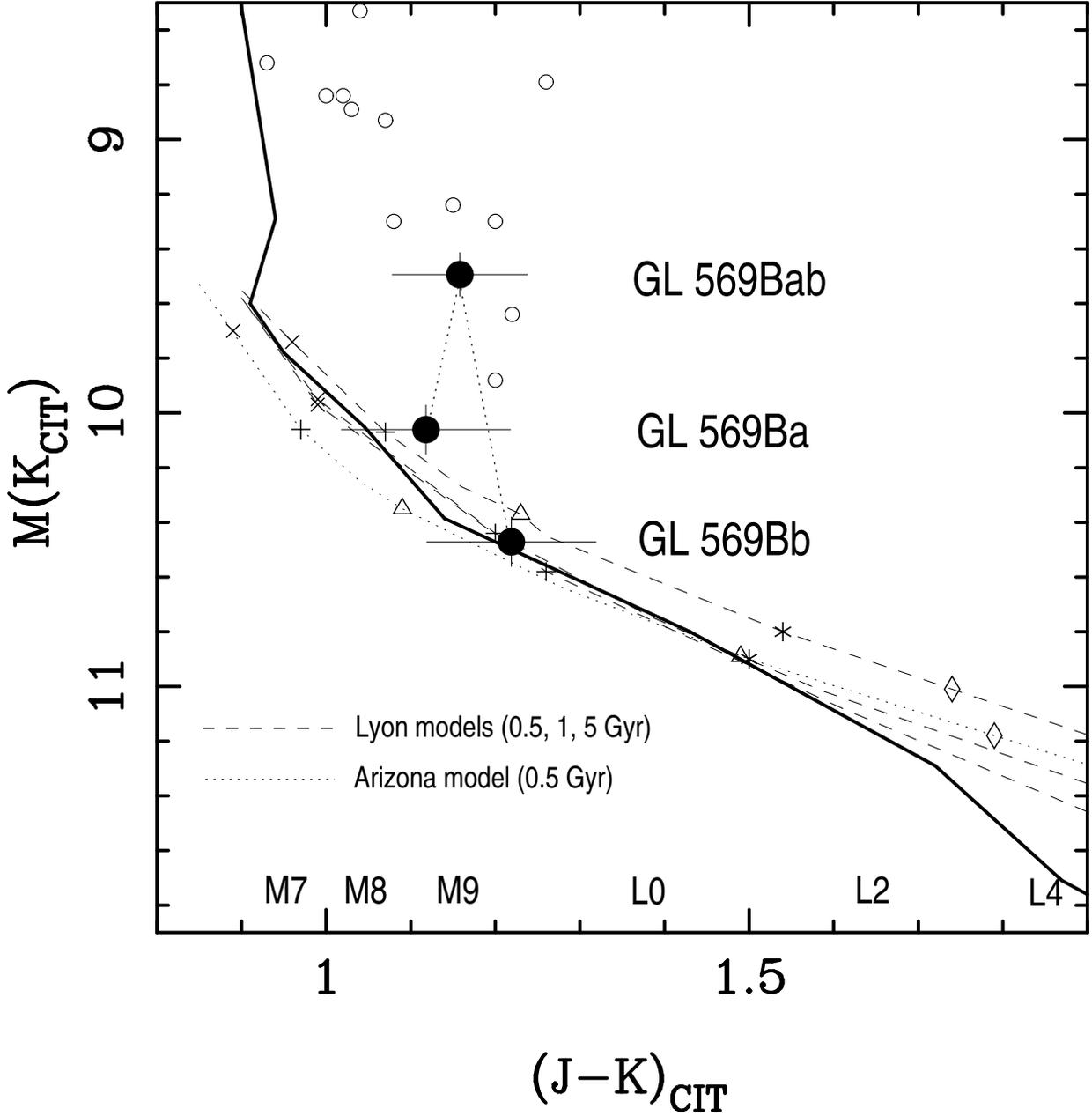}
\caption[]{\label{photfig} Infrared color-magnitude diagram displaying
  the location of Gl\,569Bab (combined light) and each of the two
  components resolved with the Keck\,II Adaptive Optics system (filled
  circles). Pleiades members ($\sim$120\,Myr) are plotted with open
  circles, and the location delineated by field dwarfs with known
  parallax is shown with a thick full line. Isochrones provided by the
  Lyon group (0.5, 1.0 and 5.0 \, Gyr, Chabrier et al$.$
  2000\nocite{chabrier00a} --- dashed lines) and by the Arizona group
  (0.5\,Gyr, Burrows et al$.$ 1997\nocite{burrows97} --- dotted line)
  are also overplotted in the diagram. Masses of 0.09\,$M_{\odot}$
  (crosses), 0.08\,$M_{\odot}$ (plus-signs), 0.072\,$M_{\odot}$ (open
  triangles), 0.060\,$M_{\odot}$ (asterisks) and 0.055\,$M_{\odot}$
  (diamonds) are marked with crosses on the isochrones. We indicate spectral
  types as a function of the ($J-K$) color on the bottom of the
  figure.}
\end{figure}

Figure~\ref{photfig} depicts the location of the Gl\,569Bab pair in
the near-infrared color-magnitude diagram. To convert their observed
magnitudes into absolute magnitudes we have used the astrometric
parallax provided by Hipparcos (0.\arcsec10191\,$\pm$\,0.\arcsec00167,
Perryman et al$.$ 1997\nocite{perryman97}), which is very similar to
previous astrometric measurements (Heintz 1991\nocite{heintz91}). Also
shown in this figure are the locations of very late-type dwarfs in the
Pleiades cluster ($\sim$120\,Myr, Basri et al$.$ 1996\nocite{basri96};
Mart\'\i n et al$.$ 1998\nocite{martin98}; Stauffer, Schultz \&
Kirkpatrick 1998\nocite{stauffer98}, we use a distance of 120 pc)
which have photometry available in the literature (Festin
1998\nocite{festin98}; Mart\'\i n et al$.$ 2000b\nocite{martin00b}),
and of objects in the field. Absolute magnitudes and colors of
M-type field standard stars have been taken from tables published in
Kirkpatrick \& McCarthy (1994)\nocite{kirk94} and Leggett et al$.$
(1998)\nocite{leggett98}.  For L-type field dwarfs we have adopted the
average near-infrared colors provided in Kirkpatrick et al$.$
(2000)\nocite{kirk00}, and have averaged absolute $K$ magnitudes for
those objects with parallax available in the literature (see Reid et
al$.$ 2000, 2001b\nocite{reid00}\nocite{reid01b}; Kirkpatrick et al$.$
2000\nocite{kirk00}).

Overplotted onto the observed data in Fig.~\ref{photfig} are the
0.5, 1.0 and 5.0 Gyr theoretical solar composition isochrones from
the evolutionary models of the Lyon group (Chabrier et al$.$
2000\nocite{chabrier00a}; Baraffe et al$.$ 1998\nocite{baraffe98}) and
from those of the Arizona group (Burrows et al$.$
1997\nocite{burrows97}).  We have adopted solar metallicity in our
studies because the photospheric abundance of the bright star Gl\,569A
has been determined to be very close to solar ([Fe/H]\,=\,--0.15,
Zboril \& Byrne 1998\nocite{zboril98}).  Although the Lyon models do
provide magnitudes and colors in the filters of interest, we preferred
to compute them from the predicted luminosity and effective
temperature at a given mass and age. This allows a direct comparison
of the two sets of interior models and minimizes the effects of
possible errors in the model atmosphere synthesis. We converted
effective temperatures ($T_{\rm eff}$) into colors by using the
temperature scales of Leggett et al$.$ (1998)\nocite{leggett98} for
M-dwarfs and of Basri et al$.$ (2000)\nocite{basri00} for late-M and
L-dwarfs. These two temperature scales should be consistent and
complementary with each other as the authors make use of the same
atmosphere models to obtain their results. We derived absolute $K$
magnitudes from theoretical luminosities by using the bolometric
correction as a function of spectral type (i.e$.$ color, $T_{\rm
  eff}$) given in Leggett et al$.$ (2000)\nocite{leggett00} down to
mid-M classes, and in Reid et al$.$ (2001b)\nocite{reid01b} for cooler
types. To summarize, the second order polynomial fits
(1700\,$\le$\,$T_{\rm eff}$\,$\le$\,3500\,K) we used are as follows:

$(J-K) = 6.423 - 3.49\times10^{-3} T_{\rm eff} + 5.41\times10^{-7} T_{\rm eff}^{2}$ \hfil $rms = 0.04$\,mag\\

$BC_K = 5.745 - 1.46\times10^{-3} T_{\rm eff} + 1.57\times10^{-7} T_{\rm eff}^{2}$ \hfil $rms = 0.06$\,mag

\noindent
Isochrones in Fig.~\ref{photfig} are plotted for $T_{\rm
  eff}$\,$\le$\,2900\,K, which roughly corresponds to masses smaller
than 0.1\,$M_{\odot}$ at ages around 1\,Gyr.

From Fig.~\ref{photfig} we can see that the evolutionary models nicely
reproduce the trend delineated by field objects, except for the
reddest colors ($J-K$\,$>$\,1.5) where models apparently predict
brighter magnitudes. Of the two sets of isochrones, the Lyon 1--5\,Gyr
models seem to produce a better fit to the observed data in the field.
The difference in color between Gl\,569Ba and Gl\,569Bb is consistent
with the spectral types of the objects. Within 1\,$\sigma$ the
uncertainties of our $JK$ photometry, the location of the pair is well
matched by isochrones in the age interval 0.2--1.0\,Gyr. This
indicates a young age for the multiple system, a result which is
compatible with the elevated X-ray emission of the ``single''
M2.5-type primary (Pallavicini, Tagliaferri \& Stella
1990\nocite{pallavicini90}; Huensch et al$.$ 1999\nocite{huensch99}),
with the system belonging to the young galactic disk as inferred from
its kinematics (Reid, Hawley \& Gizis 1995\nocite{reid95}), with the
large rotation rate measured for the star Gl\,569A (Marcy \& Chen
1992\nocite{marcy92}), as well as with the late-M spectral
type--lithium--age relationships (Magazz\`u et al$.$
1993\nocite{magazzu93}; see Bildsten et al$.$
1997\nocite{bildsten97}).

 Lithium is detected in M8--M9 Pleiades BDs (Rebolo et al$.$
1996\nocite{rebolo96}; Stauffer et al$.$ 1998\nocite{stauffer98}),
whereas older and slightly more massive objects have depleted it very
efficiently. Thus, lithium non-detections in very late-M type objects
necessarily imply ages older than the Pleiades. No lithium feature is
observed in the composite optical spectrum of Gl\,569Bab (Magazz\`u et
al$.$ 1993\nocite{magazzu93}), thus implying that the binary is older
than 0.12\,Gyr. The $R$-band spectroscopic data shown in Magazz\`u et
al$.$ (1993)\nocite{magazzu93} have poor signal-to-noise ratio, and
are dominated by the bright and more massive component as it
contributes twice as much flux as does the fainter component. We will
discuss later how the age of the system can be constrained to a much
smaller range using results presented in this paper.

We do not find from our near-infrared photometry strong evidence for
the possible binary nature of Gl\,569Ba as claimed by Mart\'\i n et
al$.$ (2000a)\nocite{martin00a} on the basis of their $H$ vs $H-K$
color-magnitude diagram as well as by Kenworthy et al $.$
(2001)\nocite{ken01} on the basis of their $J-K$ colors. The relative
position of the two components of the pair in Fig.~\ref{photfig}
reasonably fits the location of field dwarfs even within 1\,$\sigma$
the error bars. If Gl\,569Ba is a binary itself, the smaller companion
has to be at least a factor five less luminous.  We have combined our
AO $K$-band images to look for any possible companion. We place a
3\,$\sigma$ limit at $K$\,=\,16.5\,mag (0.015--0.02\,$M_{\odot}$) on
the brightness of a possible companion at distances greater than
0.25\arcsec, and less than 2\arcsec-- half the size of the AO
detector. We cannot discard, however, the presence of extremely faint
and less massive objects around any of the components which our AO
observations have not been able to detect/resolve. Follow-up high
resolution spectroscopy and/or very detailed analysis of the orbital
motion of the pair may reveal the presence of close-in giant planets.

\subsection{The HR diagram and substellarity}
\begin{figure}
\plotone{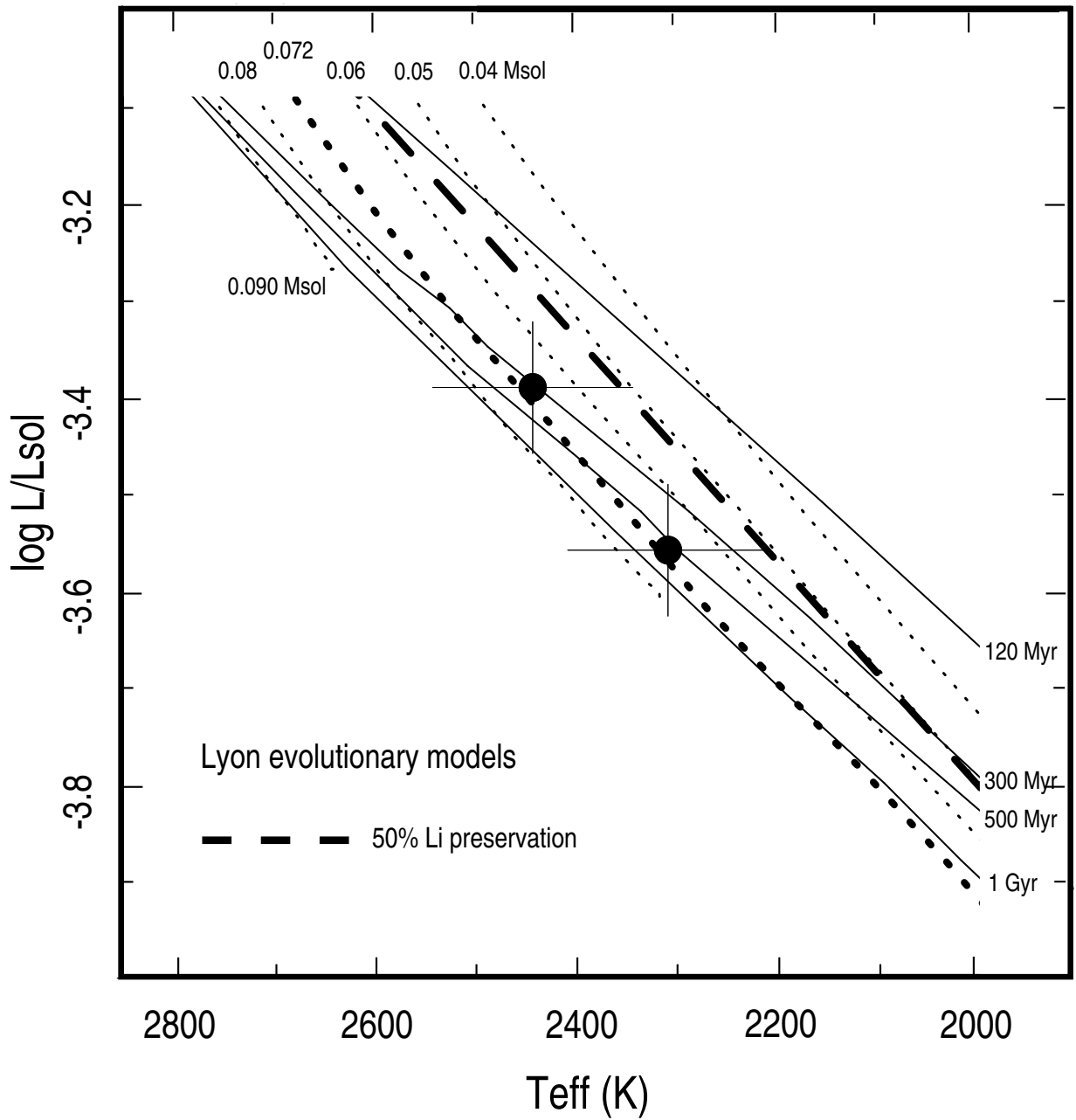}
\caption[]{\label{hrlyon} HR diagram illustrating the location of
  Gl\,569Ba and Gl\,569Bb (filled dots) in comparison with theoretical
  evolutionary tracks of constant mass (dotted lines) and isochrones
  (full lines) from the Lyon group (Chabrier et al$.$
  2000\nocite{chabrier00a}). The track corresponding to the substellar
  mass limit at 0.072\,$M_{\odot}$ is shown with a thicker dotted
  line. Masses in solar units are labelled on the upper part of the
  diagram, and ages for the isochrones are indicated to the right. The
  thick dashed line indicates the 50\%~lithium depletion limit
  predicted by the Lyon models. Objects to the left have severely
  depleted lithium, whereas objects to the right still preserve a
  significant amount of this element. Solar abundance has been assumed
  in generating this figure.}
\end{figure}

\begin{figure}
\plotone{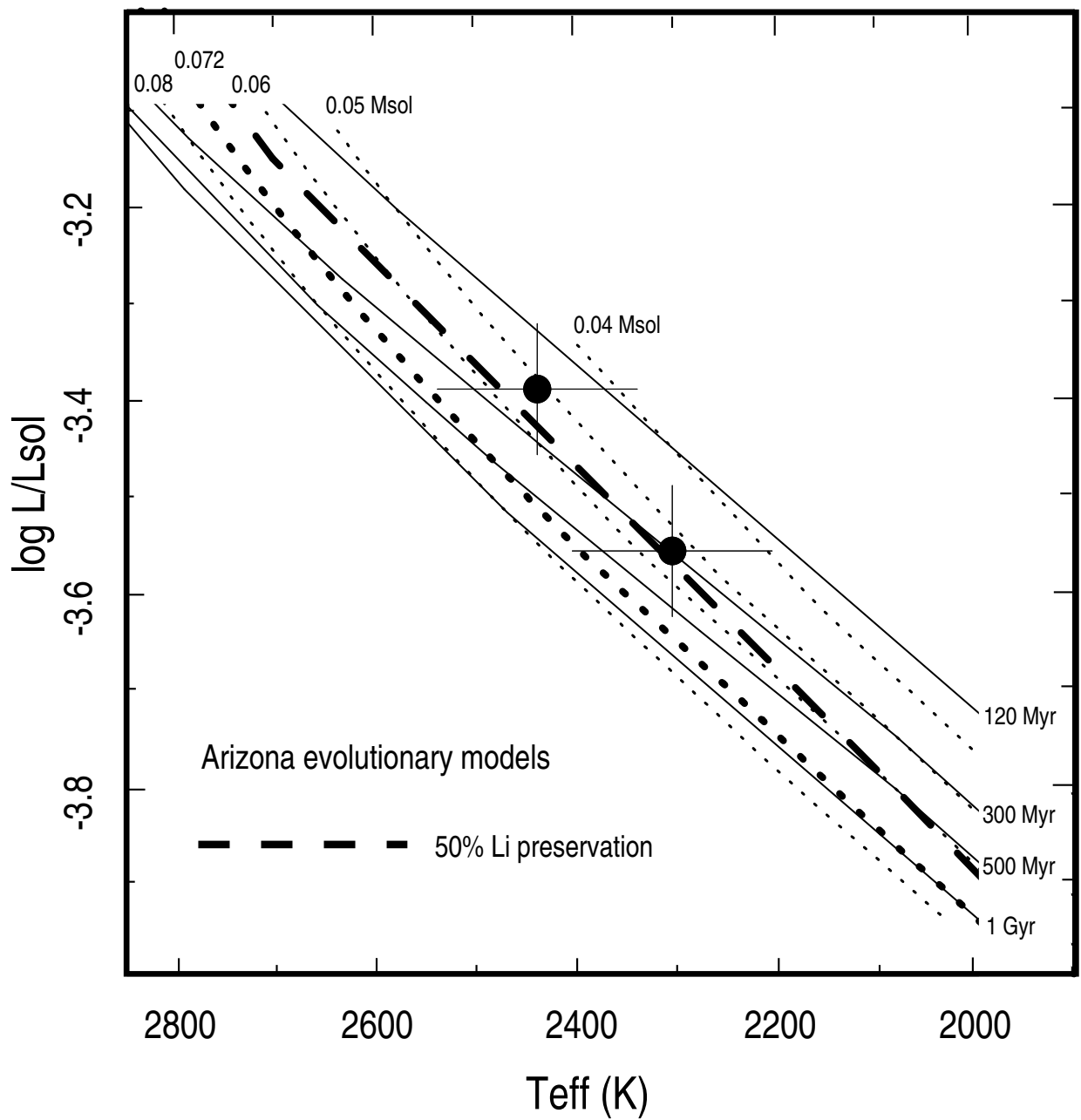}
\caption[]{\label{hrarizona} HR diagram illustrating the location of
  Gl\,569Ba and Gl\,569Bb (filled dots) in comparison with theoretical
  evolutionary tracks of constant mass (dotted lines) and isochrones
  (full lines) from the Arizona group (Burrows et al$.$
  1997\nocite{burrows97}). See the caption of Fig.~\ref{hrlyon} for
  further details. Here, the 50\%~lithium preservation line (thick
  dashed line) is taken from the Arizona models for consistency.}
\end{figure}

Figures~\ref{hrlyon} and~\ref{hrarizona} show the location of
Gl\,569Ba and Gl\,569Bb in the HR diagram (luminosity as a function of
effective temperature) and provide a comparison with state-of-the-art
evolutionary models by the Lyon group (Chabrier et al$.$
2000\nocite{chabrier00a}) and by the Arizona group (Burrows et al$.$
1997\nocite{burrows97}).  Solar-metallicity abundance isochrones of
ages 120, 300, 500\,Myr and 1\,Gyr, and evolutionary tracks of masses
in the interval 0.04--0.09\,$M_{\odot}$ are shown in these figures.
We use the most recent determination of the substellar mass limit at
0.072\,\,$M_{\odot}$ to define the stellar-substellar borderline.
Because this value is the smallest of those available in the
literature (see e.g$.$ Grossman, Hays \& Graboske
1974\nocite{grossman74}; D'Antona \& Mazzitelli
1994\nocite{dantona94}; Burrows et al$.$ 1993\nocite{burrows93}), our
conclusions on substellarity will be conservative. We indicate in the
figures the substellar mass boundary and the location of the
50\%~depletion limit of lithium burning predicted by the two sets of
models. We have obtained the luminosity and effective temperatures of
our targets as explained above. The third-order polynomial fit that
gives temperatures as a function of the observed ($J-K$) color is the
following:

$T_{\rm eff} = 7744.6 -9488.4 (J-K) + 5509.9 (J-K)^{2} - 1135.7 (J-K)^{3}$ \hfil $rms = 50$\,K

\noindent
This fit has been calculated for colors in the range 0.85--2.06
(spectral types M6--L6), and is based on the temperature calibrations
provided by Basri et al$.$ (2000)\nocite{basri00} and Leggett et al$.$
(2000)\nocite{leggett00}. Bolometric corrections in the $J$- and
$K$-bands from Reid et al$.$ (2001b)\nocite{reid01b} and Bessell,
Castelli \& Plez (1998)\nocite{bessell98}, respectively, have been
used to transform magnitudes into bolometric luminosities. The values
we derive for the pair are listed in Table~\ref{phot}. The error bars
in luminosity take into account the uncertainty of the distance
modulus (Hipparcos) and the photometric uncertainties, leading to a
total uncertainty of $\pm$0.07\,dex. The error bars assigned to the
effective temperatures come from the uncertainty in the colors alone.
These are in general a factor 2 larger than the $rms$ of the
polynomial fit describing the temperature calibration.

Models should be able to provide explanations to all physical
properties so far known for Gl\,569Ba and Gl\,569Bb, i.e., photometry,
the total mass of the pair and the destruction of lithium. From
Fig.~\ref{hrlyon} we observe that the location of Gl\,569Ba is
consistent with severe lithium depletion, in agreement with available
optical spectroscopic observations. According to the Lyon models, even
the fainter companion Gl\,569Bb has depleted its lithium. The likely
age of the system is in the range 0.2--1\,Gyr, but only with younger
ages is it possible to reproduce the astrometric total mass derived
for the pair. Therefore, the real constraint to the age of the system
is given by the total mass rather than by the error bars in the
figure. For the age of 300\,Myr, the binary would be formed by objects
of 0.069\,$M_{\odot}$ (Gl\,569Ba) and 0.059\,$M_{\odot}$ (Gl\,569Bb),
in good agreement with the mean orbital solution. Both masses are
below the stellar-substellar borderline and thus the two components
would be brown dwarfs. For slightly older ages, individual masses
would become larger, as would the total mass. On the basis of the
largest possible astrometric total mass value at the 3\,$\sigma$
level, the pair would be made up of an object on the substellar
borderline with 0.078\,$M_{\odot}$ and a 0.070\,$M_{\odot}$-BD at the
age of 500\,Myr.  We tabulate the possible masses of the pair as a
function of age in Table~\ref{masses}; these estimations rely on the
Lyon models.

The comparison with the Arizona models shown in Fig.~\ref{hrarizona}
also yields very young ages (120--500\,Myr) for the system, and thus
also small masses for each of the components. To this point, theory
and some observations seem to be consistent. However, the Arizona
evolutionary models predict hotter temperatures (by about 100\,K)
around the substellar mass limit at a given age than do the Lyon
models, and hence according to these models both Gl\,569Ba and
Gl\,569Bb should have preserved a considerable amount of lithium in
their atmospheres. Only by assuming the highest temperatures allowed
by the error bars can the Arizona models account for the observed
lithium depletion in Gl\,569Bab. This would move the pair to a
location between 300\,Myr and 500\, Myr in Fig.~\ref{hrarizona}. In
order to be consistent with the additional restriction of the
astrometric total mass, ages in the interval 300--500\,Myr are
required; the resulting masses are 0.055--0.075\,$M_{\odot}$ for
Gl\,569Ba and of 0.048--0.068\,$M_{\odot}$ for Gl\,569Bb.

Collecting evidence from the orbital solution, photometry,
spectroscopy, and the comparison with evolutionary models, the most
likely scenario of the binary Gl\,569Bab is: two very late M-type BDs
with masses of 0.055--0.078\,$M_{\odot}$ and 0.048--0.070\,$M_{\odot}$
in a close orbit, in turn orbiting an early M-type 0.5\,$M_{\odot}$
star, the whole system with an age in the range 250--500\,Myr. Such
young ages found in nearby objects are not surprising since it now
seems that the Sun is located close to a region that was the site of
substantial amounts of recent stellar formation (Zuckerman \& Webb
2000\nocite{zuckerman00}).

\begin{figure}
\epsscale{1.0}
\plotone{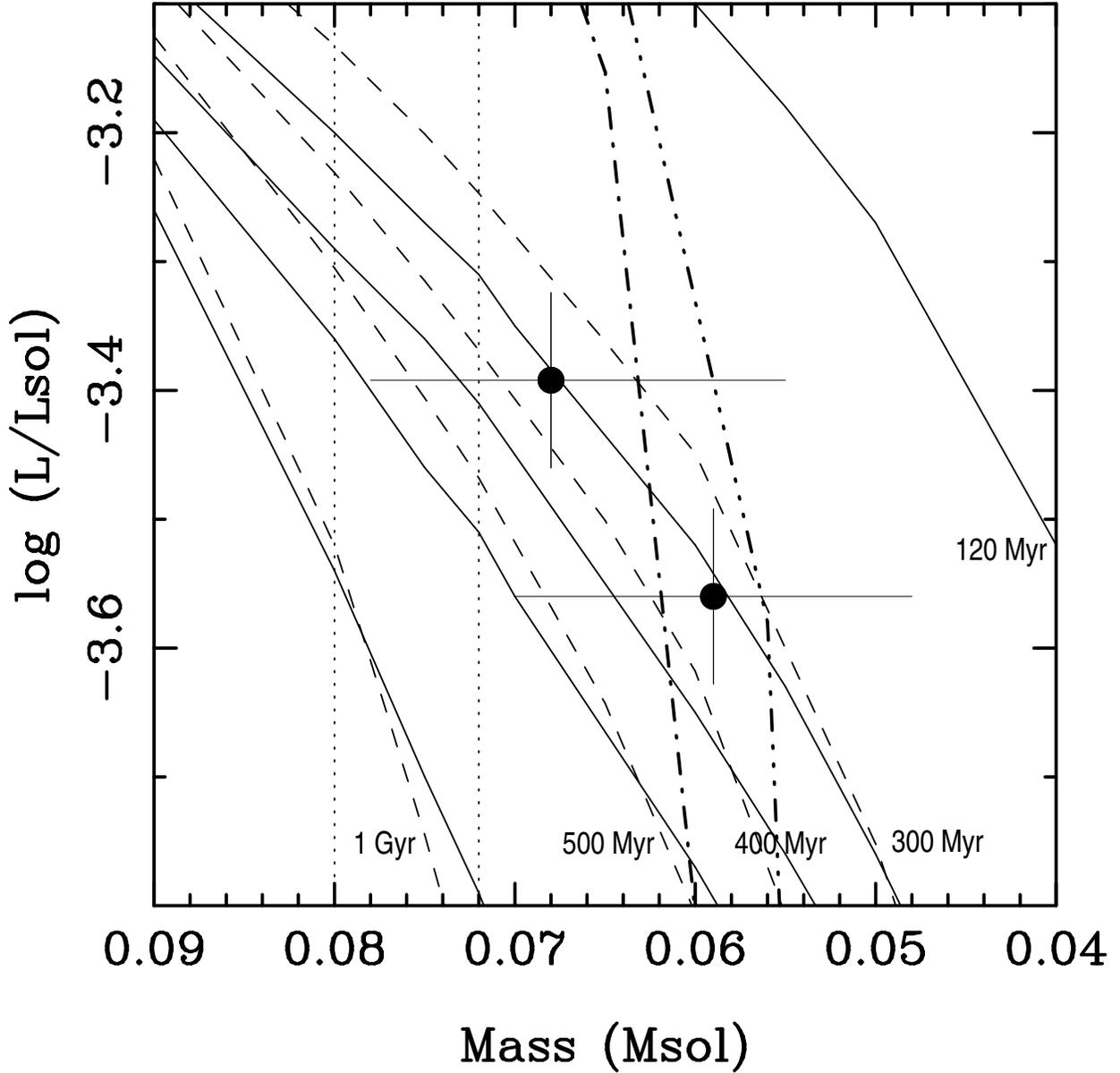}
\caption[]{\label{masslum} Mass-luminosity relationship for different
  ages (120, 300, 400 and 500\,Myr, and 1\,Gyr) according to the Lyon
  models (Chabrier et al$.$ 2000\nocite{chabrier00a} --- full lines)
  and to the Arizona models (Burrows et al$.$ 1997\nocite{burrows97}
  --- dashed lines). The stellar-substellar transition mass range at 0.072--0.080\,$M_{\odot}$
  is indicated by the vertical dotted lines. The dot-dot-dot-dashed
  line (Lyon) and the dot-dashed line (Arizona) mark when 50\%~of the
  lithium has burned. Lithium preservation occurs to the right of
  these lines. The error bars assigned to Gl\,569Ba and Gl\,569Bb
  (filled circles) correspond to the uncertainty in luminosity, and to
  the likely mass range (3-$\sigma$) derived for each object.}
\end{figure}

Figure~\ref{masslum} portrays the mass-luminosity relationship for
different ages as given by the evolutionary models. The two members of
the pair are plotted with error bars indicating the uncertainty in
luminosity and the likely mass range of each component. Masses that
have depleted lithium by a factor 2 are also incorporated into the
figure. According to the Arizona models, Gl\,569Bb may have preserved
lithium in its atmosphere, whereas this is quite unlikely based on the
Lyon models.  Therefore, lithium observations of this BD are needed in
order to discriminate which model reproduces neatly the properties of
the pair.  In addition, precise radial velocity measurements of each
component will lead to an accurate determination of the individual
masses, and thus will also constitute a better constraint on the
models.  Nevertheless, models do not appear to be far from reproducing
the observational properties of Gl\,569Bab. Delfosse et al$.$
(2000)\nocite{delfosse00} show that the mass-luminosity relationship
given by the Lyon models reasonably describes the low-mass stellar
regime in the field. The pair Gl\,569Bab has lower masses that belong
to the substellar regime. Brown dwarfs around stars have been reported
in the recent years (see Table~5 in Reid et al$.$
2001a\nocite{reid01a} for a compilation of the complete list), but to
our knowledge none of them has been proved to be a binary itself.
Gl\,569Bab turns out to be the first confirmed resolved binary BD as a
companion to a star.

The distance to Gl\,569 implies a physical separation of 49\,AU
between the M2.5-type star and the substellar pair. This large
separation and the high mass ratio ($q$\,$\sim$\,0.135 and 0.120)
between the star and each of the BDs favors the fragmentation of a
self-gravitational collapsing molecular cloud as the most plausible
explanation for the formation of the system (Boss 2000\nocite{boss00};
Bodenheimer 1998\nocite{bodenheimer98}). Whether each component of the
pair Gl\,569Bab originated from a second fragmentation and collapse
process of a small cloud core is not clear (the physical separation is
0.92\,AU, and the mass ratio of the pair is $q$\,$\sim$\,0.89). The
activity of the nascent low-mass star when it was gaining mass and
becoming more luminous may have caused the disruption of the less
massive collapsing core into two close substellar objects before the
hydrostatic core could build up enough mass to eventually start
hydrogen burning. Energetic outflows and jets up to thousands of AU in
length have been detected in low-mass stars of very young star forming
regions (e.g., Reipurth et al$.$ 2000\nocite{reipurth00}; Fridlund \&
Liseau 1998\nocite{fridlund98}).  We cannot discard the possibility,
however, that the protoplanetary disk around the star might have also
played an important role in the origin of the companions. Disks
extending up to several hundred AU are known to exist around stars
(Bruhweiler et al$.$ 1997\nocite{bruhweiler97}).  Clearly, finding
other similar systems will, in addition to providing additional
dynamical masses, also contribute to our knowledge of the genesis of
such interesting multiple low-mass binaries.

\section{Conclusions}
We have obtained new observations of the Gl\,569Bab pair (low-mass
binary companion at a wide separation from the M2.5-type star
Gl\,569A), which have allowed us to derive the spectral types of each
component and the orbital parameters of the system. We find that the
total mass of the low-mass binary is $0.123_{-0.022}^{+0.027}
M_{\odot}$ (3-$\sigma$) with two detected components of M8.5 and M9
spectral types (half a subclass uncertainty) completing one eccentric
($e$\,=\,0.32\,$\pm$\,0.02) orbit every 892\,$\pm$25\,days. We have
also acquired new $J$ and $K$ near-IR photometry in order to locate
Gl\,569Ba and Gl\,569Bb in the HR diagram and compare them with the
most recent evolutionary models by the Lyon group (Chabrier et al$.$
2000\nocite{chabrier00a}) and the Arizona group (Burrows et al$.$
1997\nocite{burrows97}). The pair is likely formed by two solar
metallicity young brown dwarfs with masses in the interval
0.055--0.078\,$M_{\odot}$ (Gl\,569Ba) and 0.048--0.070\,$M_{\odot}$
(Gl\,569Bb) at the young ages of 250--500\,Myr. Our adaptive optics
images taken with the Keck\,II telescope exclude the presence of any
other resolved companion with $K$ magnitudes brighter than 16.5
(3\,$\sigma$) at separations of 0.25\arcsec~up to 2\arcsec~from
Gl\,569Bab. This detection limit corresponds to masses around
0.015--0.02\,$M_{\odot}$ for the possible age range of the system.
Further radial velocity and astrometric measurements will be very
valuable to detect giant planets, as well as to provide individual
masses for each of the members of the pair.

\acknowledgments 

We thank N. Reid for valuable comments and discussion. Data presented
herein were obtained at the W.\,M.  Keck Observatory, which is
operated as a scientific partnership among the California Institute of
Technology, the University of California and the National Aeronautics
and Space Administration (the Observatory was made possible by the
generous financial support of the W.\,M. Keck Foundation); and at the
Carlos S\'anchez Telescope (CST) operated on the island of Tenerife in
the Spanish Observatorio del Teide of the Instituto de Astrof\'\i sica
de Canarias. This research has made use of the Simbad database,
operated at CDS, Strasbourg, France. We are thankful to I. Baraffe and
A. Burrows for providing us computer-readable files of their
evolutionary models.

\clearpage

\clearpage

\clearpage

\clearpage

\clearpage
\clearpage

\clearpage

\clearpage

\clearpage 

\begin{table}
\begin{center}
\caption[]{\label{phot} Photometry (CIT system) and physical parameters of Gl\,569Ba and Gl\,569Bb}
\begin{tabular}{lcccccc}
           &        &       &       &        &         \\
\tableline\tableline
\multicolumn{1}{c}{Object}     &  
\multicolumn{1}{c}{$J$}     &  
\multicolumn{1}{c}{$K$}     &  
\multicolumn{1}{c}{$J-K$}    &
\multicolumn{1}{c}{log\,$L/L_{\odot}$}  &
\multicolumn{1}{c}{$T_{\rm eff}$ (K)}  &
\multicolumn{1}{c}{$M/M_{\odot}$}  \\
\tableline
Gl\,569Bab & 10.61\,$\pm$\,0.05 &  9.45\,$\pm$\,0.05 & 1.16\,$\pm$\,0.07 & --3.17\,$\pm$\,0.07 &                  & 0.101--0.150\\
Gl\,569Ba  & 11.14\,$\pm$\,0.07 & 10.02\,$\pm$\,0.08 & 1.12\,$\pm$\,0.10 & --3.39\,$\pm$\,0.07 & 2440\,$\pm$\,100 & 0.055--0.078\\
Gl\,569Bb  & 11.65\,$\pm$\,0.07 & 10.43\,$\pm$\,0.08 & 1.22\,$\pm$\,0.10 & --3.56\,$\pm$\,0.07 & 2305\,$\pm$\,100 & 0.048--0.070\\
\tableline
\end{tabular}
\tablecomments{The relative photometry between Gl\,569Ba and Gl\,569Bb is known to a better accuracy (see text).}
\end{center}
\end{table}

\begin{table}
\begin{center}
\caption[]{\label{ast} Astrometry of Gl\,569Ba-Bb}
\begin{tabular}{cccc}
             &           &                    &       \\
\tableline\tableline
   Date      & Epoch     & Separation         & P.A.  \\
             & (MJD)     &  (\arcsec)         & ($^{\circ}$) \\
\tableline
1999 Aug 29  & 51419.270 & $0.101 \pm 0.001  $ & $46.8  \pm 3$ \\
2000 Feb 18  & 51592.600 & $0.092 \pm 0.001  $ & $98.2  \pm 3$ \\
2000 Feb 25  & 51599.644 & $0.090 \pm 0.001  $ & $100.4 \pm 2$ \\
2000 Jun 20  & 51715.408 & $0.076 \pm 0.003  $ & $138.6 \pm 2$ \\
2000 Aug 24  & 51780.283 & $0.059 \pm 0.001  $ & $178.4 \pm 2$ \\
2001 Jan 09  & 51918.665 & $0.073 \pm 0.002  $ & $291.4 \pm 2$ \\
2001 May 10  & 52039.410 & $0.097 \pm 0.001  $ & $341.1 \pm 3$\\

\tableline
\end{tabular}
\end{center}
\end{table}

\begin{table*}
\begin{center}
%\tabletypesize{\scriptsize}
\caption[]{\label{eqw} K\,{\sc i} and Na\,{\sc i} equivalent widths (\AA) and the strengths of the H$_2$O and CO bands}
\begin{tabular}{lcccccccccc}
          &      &     &     &     &     & &     &      &      &      \\
\tableline\tableline
\multicolumn{2}{c}{} &
\multicolumn{4}{c}{K\,{\sc i} wavelength ($\mu$m)} &
\multicolumn{1}{c}{} &
\multicolumn{2}{c}{Na\,{\sc i} wavelength ($\mu$m)} &
\multicolumn{1}{c}{H$_2$O\tablenotemark{a}} &
\multicolumn{1}{c}{CO\tablenotemark{b}} \\
Object   &  SpT & 1.169\tablenotemark{c} & 1.177\tablenotemark{c} & 1.244\tablenotemark{c} & 1.253\tablenotemark{c} & & 2.206\tablenotemark{c} & 2.209\tablenotemark{c} & 1.330\tablenotemark{c} & 2.294\tablenotemark{c}\\
\tableline
Gl\,569Ba & M8.5 & 6.5 & 7.5 & 4.7 & 4.8 & & 1.4 & 0.70 & 0.72 & 1.21 \\
Gl\,569Bb & M9.0 & 6.5 & 7.7 & 5.2 & 5.1 & & 0.9 & 0.50 & 0.70 & 1.24 \\
\tableline
\end{tabular}
\tablenotetext{a}{Ratio of the average flux in a 0.02\,$\mu$m window centred on 1.34\,$\mu$m and on 1.29\,$\mu$m (Reid et al$.$ 2001a).}
\tablenotetext{b}{Ratio of the average flux in a 0.06\,$\mu$m window centred on 2.25\,$\mu$m and on 2.33\,$\mu$m (Jones et al$.$ 1994).}
\tablenotetext{c}{All wavelengths in $\mu$m.}
\tablecomments {Uncertainties are $\pm$0.5 for the spectral classification, 10\%~for equivalent widths and 5\%~for the flux ratios.}
\end{center}
\end{table*}

\begin{table}
\begin{center}
\caption[]{\label{params}Orbital parameters of Gl\,569Ba-Bb}
\begin{tabular}{lrcl}
             &          \\   
\tableline\tableline
\multicolumn{1}{l}{Parameter} &
\multicolumn{3}{c}{Value} \\
\tableline
Period, $P$                       & 892&$\pm$&25 day \\
Eccentricity, $e$                 & 0.32&$\pm$&0.02 \\
Semi-Major Axis, $a$              & 0.90&$\pm$&0.02 AU \\
Inclination, $i$                  & 34&$\pm$&3 deg \\
Arg. Periapsis, $\omega$          & 76&$\pm$&4 deg \\
Long. of Ascending Node, $\Omega$ & 141&$\pm$&4 deg \\
Epoch (MJD), $T$                  & 51820&$\pm$&4 day \\
\tableline
\end{tabular}
\tablecomments {Uncertainties are 1-$\sigma$. Note that the toal mass uncertainty is smaller 
than this indicates, see Fig. \ref{chi} and text.}
\end{center}
\end{table}

\begin{table}
\begin{center}
\caption[]{\label{masses} Likely ages and individual masses of
                          Gl\,569Ba and Gl\,569Bb}
\begin{tabular}{cccc}
           & & & \\
\tableline\tableline
\multicolumn{1}{c}{Age}       & \multicolumn{1}{c}{Gl\,569Ba} & \multicolumn{1}{c}{Gl\,569Bb} & \multicolumn{1}{c}{Total mass}\\
\multicolumn{1}{c}{(Myr)}       & \multicolumn{1}{c}{($M_{\odot}$)} & \multicolumn{1}{c}{($M_{\odot}$)} & \multicolumn{1}{c}{($M_{\odot}$)}\\
\tableline
200 & 0.055 & 0.048\tablenotemark{a} & 0.103 \\
250 & 0.061 & 0.053\tablenotemark{a} & 0.114 \\
300\tablenotemark{b} & 0.069\tablenotemark{b} & 0.059\tablenotemark{b}
                     & 0.128\tablenotemark{b} \\
400 & 0.072 & 0.065 & 0.137 \\
500 & 0.078 & 0.070 & 0.148 \\
700\tablenotemark{c} & 0.082\tablenotemark{c} & 0.074\tablenotemark{c} & 0.156\tablenotemark{c} \\
\tableline
\end{tabular}
\tablenotetext{a}{Should have (partially) preserved lithium.}
\tablenotetext{b}{These estimates match the mean astrometric orbital
  solution.}
\tablenotetext{c}{The total mass inferred for this age is even
  beyond the 3\,$\sigma$ uncertainty of our astrometric solution.}
\tablecomments{Based on the Lyon models (Chabrier et al$.$
  2000\nocite{chabrier00a}; Baraffe et al$.$ 1998\nocite{baraffe98}).
  The uncertainty of individual mass models is
  $\pm$0.002\,$M_{\odot}$.}
\end{center}
\end{table}

\end{document}